\documentclass[aip,chaos,reprint]{revtex4-1}

\usepackage{amsmath}
\usepackage{amssymb}
\usepackage{graphicx}
\usepackage{color}

\begin{document}
\title{Colossal Brownian yet non-Gaussian diffusion in a periodic potential: \\impact of nonequilibrium noise amplitude statistics}
\author{K. Bia{\l}as}
\affiliation{Institute of Physics, University of Silesia, 41-500 Chorz{\'o}w, Poland}
\author{J. Spiechowicz}
\email{jakub.spiechowicz@us.edu.pl}
\affiliation{Institute of Physics, University of Silesia, 41-500 Chorz{\'o}w, Poland}

\begin{abstract}
Last year in [Phys. Rev. E 102, 042121 (2020)] the authors studied an overdamped dynamics of nonequilibrium noise driven Brownian particle dwelling in a spatially periodic potential and discovered a novel class of Brownian, yet non-Gaussian diffusion. The mean square displacement of the particle grows linearly with time and the probability density for the particle position is Gaussian, however, the corresponding distribution for the increments is non-Gaussian. The latter property induces the colossal enhancement of diffusion, significantly exceeding the well known effect of giant diffusion. 
Here we considerably extend the above predictions by investigating the influence of nonequilibrium noise amplitude statistics on the colossal Brownian, yet non-Gaussian diffusion. The tail of amplitude distribution crucially impacts both the magnitude of diffusion amplification as well as Gaussianity of the position and increments statistics. Our results carry profound consequences for diffusive behaviour in nonequilibrium settings such as living cells in which diffusion is a central transport mechanism.

\end{abstract}

\maketitle

\begin{quotation}
An equilibrium system is ruled by monumental Thermodynamic Laws and various symmetries such as, for instance, a \emph{detailed balance}. When taken out of equilibrium they generally lose their validity. Solely this fact opens a new horizon which to a large extent still remains a \emph{terra incognita} in physics. Yet, some progress has been achieved in exploring this fascinating ground. Understanding phenomena like stochastic and coherence resonance, ratchet effects, negative mobility and anomalous diffusion testifies it. Here we reveal another face of nonequilibrium. We exemplify that solely a nonequilibrium state created by an external stochastic force can serve as a seed for a diffusion anomaly in which normal, Brownian scaling of the displacement is reconciled with the non-Gaussian statistics. This finding must be contrasted with the most fundamental hallmarks of Brownian diffusion which are linear scaling of the mean square displacement with time and Gaussianity of the probability density for the particle coordinate and its increments.
\end{quotation}

\section{Introduction}
The theoretical foundations of Brownian motion were formulated by Sutherland, Einstein, Smoluchowski and Langevin \cite{sutherland,einstein,smoluchowski,langevin} at the beginning of the XX century. Ever since it has enjoyed an unfading interest \cite{kanazawa2020, neupane2016, jeon2016, spiechowicz2016njp, peng2016, kim2017, spiechowicz2016scirep, spiechowicz2017scirep, zhang2017, illien2018, goychuk2019, spiechowicz2019chaos, li2019, spiechowicz2019njp, spiechowicz2021pre_arcsine}. 
One of its signs is a novel type of diffusion processes in which a linear growth of the mean square displacement is observed, yet with a non-Gaussian distribution for the particle coordinate \cite{wang2012}. Such Brownian, yet non-Gaussian diffusion has been explained theoretically by either space- or time-dependent diffusion coefficients, reflecting the characteristic features of the particle environment described in terms of superstatistics \cite{wang2009,hapca2009} or by a diffusing diffusivity model \cite{chubynsky2014,jain2016,metzler2017,tyagi2017}. This peculiar behaviour is detected typically in biological systems, such as soft and active matter setups \cite{wang2012,goldstein2009,metzler2019,barkai2020physreve} or in a large class of transport problems in random media \cite{jl1,jl2,barkai2020}.

Last year yet another manifestation of this unusual diffusion phenomenon has been communicated where the mean square displacement scales linearly with time, the probability distribution for the particle position is Gaussian, however, the corresponding probability density for the increments deviates from Gaussian and displays an exponential tail \cite{bialas}. It is in clear contrast to Brownian motion for which the increments are described by Gaussian statistics too. The authors considered an archetypal model of a nonequilibrium system, namely, an overdamped Brownian particle dwelling in a periodic potential, which additionally was under the action of a biased nonequilibrium noise. This case is vital for understanding transport in both physical and biological systems. In the latter setups typically there is no systematic gradient or force but instead they are immersed in a sea of random perturbations. The existence of the exponential tail in the increment statistics induces the colossal enhancement of diffusion, pronouncedly exceeding the effect well known as giant diffusion \cite{constantini1999, reimann2001a, reimann2002, lindner2001, lindner2016, spiechowicz2020pre, spiechowicz2021pre_weak} which was experimentally confirmed in various setups \cite{lee2006,reimann2008,ma2015}.

In Ref. [\onlinecite{bialas}] the authors wrote \textit{"The question is how the diffusion coefficient and the probability density of the particle position increments depends on the choice of the nonthermal noise statistics... In general, it is a very complex problem and may be resolved only in a case by case manner"}. Therefore in this paper we revisit this problem to analyze the impact of nonequilibrium noise amplitude statistics on the colossal Brownian, yet non-Gaussian diffusion. We study a broad spectrum of different probability distributions that allows us to consider the influence of non-monotonicity as well as exponential, superexponential (e.g. Gaussian) and subexponential (algebraic) decay in the density. In particular, we want to address the following essential questions: (i) How does the magnitude of diffusion enhancement depend on the amplitude statistics? (ii) Is the existence of the exponential tail in the increment distribution robust with respect to the alteration of the amplitude statistics? (iii) Can the Gaussianity of the probability density for the particle position be controlled by modification of the amplitude statistics?

The work is organized in the following way. In Sec. II we recall the overdamped dynamics of the Brownian particle dwelling in a periodic potential, discuss the dimensionless units as well as the nonequilibrium noise amplitude statistics and present the diffusion coefficient as a main quantity of interest. Next, in Sec. III we elaborate on results with a particular focus on two aspects, namely, the colossal enhancement of diffusion and Brownian, yet non-Gaussian diffusion. Sec. IV provides a summary and final conclusions.

\section{Model}
We start our analysis with the overdamped Langevin dynamics for the Brownian particle moving in a periodic potential and subjected to an external force, namely
\begin{equation}
\label{dim_model}
	\Gamma \dot{x} =-U'(x) + F(t) + \sqrt{2\Gamma k_B T}\,\xi(t).
\end{equation}
The potential is postulated in the simple, spatially periodic form
\begin{equation}
	U(x) = \Delta U \sin{\left(2\pi \frac{x}{L} \right)}.
\end{equation}
The dot and prime denote differentiation with respect to time $t$ and coordinate $x$ of the Brownian particle, respectively. 
The parameter $\Gamma$ stands for the friction coefficient, $F(t)$ corresponds to an external force, $k_B$ is the Boltzmann constant and $T$ is thermostat temperature. Thermal fluctuations are modeled by $\delta$-correlated Gaussian white noise $\xi(t)$ of vanishing mean $\langle \xi(t) \rangle = 0$ and the correlation function $\langle \xi(t)\xi(s) \rangle = \delta(t-s)$.


To make Eq. (\ref{dim_model}) dimensionless we rescale the particle coordinate and time as
\begin{equation}
	\label{scales}
	 \hat{x}=\frac{2\pi}{L}x, \quad \hat{t}=\frac{t}{\tau_0}, \quad \tau_0=\frac{1}{4 \pi^2} \frac{\Gamma L^2}{\Delta U}.
\end{equation}
Note that $\tau_0$ is related to the characteristic time scale for an overdamped particle to move from the maximum of the potential $U(x)$ to its minimum. This procedure allows to simplify the description of model as after such a transformation a number of free parameters is reduced. After the above redefinition the equation is
\begin{equation}
\label{dimless_model}
	\dot{x} =-\hat{U}'(\hat{x}) + f(t) + \sqrt{2D_T}\,\hat{\xi}(\hat{t}),
\end{equation}
where the rescaled potential
\begin{equation}
	\hat{U}(\hat{x}) = \frac{1}{\Delta U} \, U\left( \frac{L}{2\pi} \hat{x} \right) = \sin{\hat{x}}
\end{equation}
has the period $2\pi$ and the barrier height $2$. The dimensionless friction coefficient $\gamma = 1$ whereas the external force acting on the particle is
\begin{equation}
	f(\hat{t}) = \frac{1}{2\pi}\frac{L}{\Delta U} F(\tau_0 \hat{t}).
\end{equation}
Likewise, the dimensionless thermal noise takes the form
\begin{equation}
	\hat{\xi}(\hat{t}) = \frac{1}{2\pi} \frac{L}{\Delta U}\, \xi(\tau_0 \hat{t})
\end{equation}
and possesses the same statistical properties as $\xi(t)$, i.e. it is Gaussian stochastic process with the vanishing mean $\langle \hat{\xi}(\hat{t}) \rangle = 0$ and the correlation function $\langle \hat{\xi}(\hat{t})\hat{\xi}(\hat{s})\rangle = \delta(\hat{t} - \hat{s})$. The dimensionless thermal noise intensity $D_T = k_B T/\Delta U$ is a ratio of the thermal and half of the activation energy the particle needs to overcome the non-rescaled potential barrier. Hereafter, we use only the dimensionless quantities and we therefore omit the $\wedge$-notation to simplify the look of equations.

In this work we analyze two variants of the external force $f(t)$ appearing in the archetypal model of a nonequilibrium system given by Eq. (\ref{dimless_model}). First, we consider a constant bias $f(t) = f$. It is known that for such a setup at a critical force $f = f_c$ the diffusion coefficient pronouncedly surpasses the free diffusion coefficient. This effect was captured as the giant diffusion \cite{reimann2001a,reimann2002}. However, in many real systems, such as e.g. living cells \cite{bressloff2013}, for a strongly fluctuating environment there is no systematic deterministic load but rather random collisions or releases of chemical energy. Therefore as the second variant of the external force $f(t)$ we consider a biased nonequilibrium noise $f(t) = \eta(t)$ which pumps energy to the system in a non-deterministic way. To compare these both scenarios we fix the mean value of the nonequilibrium noise $\eta(t)$ equal to the bias $f$, i.e. $\langle \eta(t) \rangle = f$.

To model such a perturbation we employ the stochastic biased force $\eta(t)$ in the form of a sequence of $\delta$-shaped pulses with random amplitudes $z_i$ defined in terms of the biased white Poissonian shot noise \cite{spiechowicz2013jstatmech,hanggi1978,hanggi1980}, namely
\begin{equation}
    \eta(t)=\sum^{n(t)}_{i=1}z_i\delta (t-t_i),
\end{equation}
where $t_i$ are counting times of a Poissonian process $n(t)$ described by the parameter $\lambda$. The probability distribution for occurrence of $k$ impulses in the time interval $[0,t]$ is given in the Poisson form \cite{feller1970}
\begin{equation}
	Pr\{ n(t) = k \} = \frac{(\lambda t)^k}{k!} e^{-\lambda t}.
\end{equation}
The amplitudes $\left\{ z_i \right\}$ are independent random variables sampled from a probability distribution $\rho\left(z\right)$. In this work we consider four different classes of the latter function, i.e. exponential, half-normal, Erlang and Lomax distribution. All of them are multiplied by the Heaviside step function $\theta(z)$ and as a consequence amplitudes $\{z_i\}$ are positive. Their mean value reads $\langle z_i \rangle = \zeta$. As a result realizations of the stochastic biasing force are \emph{non-negative}, i.e. $\eta(t) \ge 0$.

For all of the above distributions $\eta(t)$ embodies white noise of a finite mean and a covariance given by \cite{spiechowicz2014pre}
\begin{subequations}
\begin{align}
	\langle \eta(t) \rangle &= \lambda \langle z_i \rangle = \lambda \zeta, \\
	\langle \eta(t)\eta(s) \rangle - \langle \eta(t) \rangle \langle \eta(s) \rangle &= 2 D_P \delta (t-s),
\end{align}
\end{subequations}
where we introduced the Poissonian shot noise intensity $D_P = \lambda \langle z_i^2 \rangle/2$. We also assume that thermal fluctuations $\xi(t)$ are uncorrelated with nonequilibrium noise $\eta(t)$, i.e. $\langle \xi(t) \eta(s) \rangle = \langle \xi(t) \rangle \langle \eta(s) \rangle = 0$. 
Parameter $\lambda$ may be interpreted as the average frequency of the $\delta-$spikes whereas $\zeta$ is the mean amplitude of the single pulse. It means that, for instance, if $\lambda$ is large and $\zeta$ small, then the particle is frequently kicked by small impulses. On the other hand, if $\lambda$ is small and $\zeta$ large then it is rarely kicked by large $\delta$-spikes.
\begin{figure}[t]
	\centering
    \includegraphics[width=0.9\linewidth]{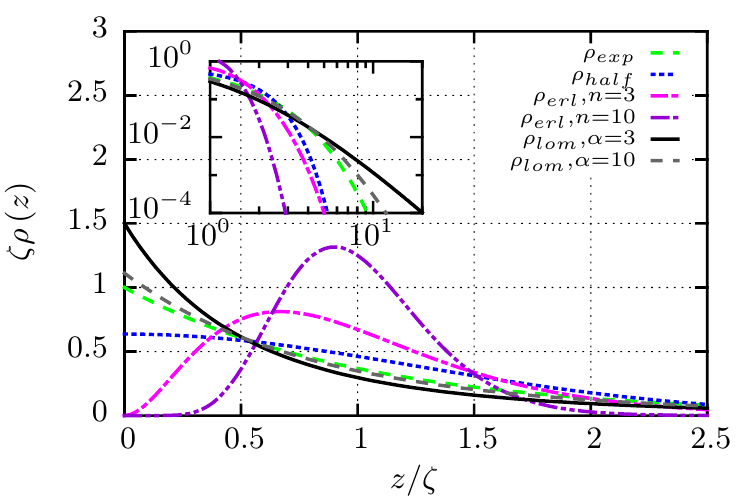}
    \caption{The comparison of the probability densities $\rho(z)$ of the amplitudes  $\{z_i\}$ considered in this work. Note that only the Erlang distribution displays a non-monotonic function. The Lomax (Pareto) density is an example of a fat-tailed distribution, see the inset.}
    \label{fig1}
\end{figure}

\subsection{Poissonian noise amplitude distributions}
In this subsection we describe all amplitude distributions which will be taken into account in our study. 

\subsubsection{Exponential distribution}
Exponential distribution is found in plentiful of different contexts, in particular when describing the lengths of inter-arrival times in a homogeneous Poisson counting process. It can be parameterized as follows
\begin{equation}
    \rho_{exp}(z) =  \frac{\theta(z)}{\mu} \mbox{exp}\left(-\frac{z}{\mu}\right)
    \label{pdf_exp}
\end{equation}
where it is characterized by only one parameter $\mu > 0$ being equal to the mean $\zeta = \mu$. The variance of the distribution reads $\sigma^2 = \mu^2$. Exponential distribution is the only continuous distribution that is memoryless \cite{feller1970} implying that the distribution of a waiting time until certain event does not depend on how much time has elapsed. This distribution also serves as a benchmark for how fast distribution vanishes. Those that have heavier tails (are not exponentially bounded) are called heavy-tailed distributions \cite{rolski1999}. To avoid confusion of the reader we stress that the amplitude probability distribution $\rho\left(z\right)$ is independent of  the distribution of time $s = t_i - t_{i-1}$ between successive Poisson arrival times which is likewise exponential.

\subsubsection{Half-normal distribution}
Half-normal distribution \cite{johnson1995} has the form
\begin{equation}
    \rho_{half}(z) = \frac{ \theta(z)\sqrt{2}}{\mu \sqrt{\pi}}\mbox{exp}\left(-\frac{z^2}{2\mu^2}\right)
    \label{pdf_half}
\end{equation}
with the scale parameter $\mu>0$. Its mean is $\zeta=\mu\sqrt{2}/\sqrt{\pi}$ and the variance $\sigma^2=\mu^2\left(1-2/\pi\right)$. If $X$ follows a normal distribution with zero mean and the variance $\mu^2$ then $\left|X\right|$ is a half-normal distributed random variable, hence its name. It vanishes faster the than exponential distribution and therefore does not represent the class of heavy-tailed densities. Some applications of the half-normal distribution include e.g. modeling the lifetime data \cite{cooray2008}. 

\subsubsection{Erlang distribution}
Erlang distribution \cite{johnson1995} is a generalization of exponential distribution. Its probability density function is
\begin{equation}
    \rho_{erl}(z,n) = \frac{\theta(z) z^{n-1}}{{\mu}^n(n-1)!}  \mbox{exp}\left (-\frac{z}{\mu} \right),
    \label{pdf_erlang}
\end{equation}
where $n > 0$ is a positive integer shape parameter and $\mu>0$. This distribution has a mean $\zeta=n\mu$ and variance $\sigma^2= n \mu^2$. It is a probability density of a sum of $n$ independent exponential variables with mean $\mu$. Consequently, in the context of Poisson point process Erlang distribution describes distribution of a time interval which elapsed up to the $n$-th event. In contrast to previous distributions it is an example of the non-monotonic probability density. For $n>1$ it possesses a maximum at $z = (n-1)\mu > 0$ unlike other variants considered in this study for which the extremum is observed at $z = 0$. As it can be seen in Fig. \ref{fig1} when $n$ is increased distribution becomes more localized if its mean is fixed. From this observation we can conclude that its tail vanishes faster for larger $n$. Curiously, the age distribution of cancer incidence often follows the Erlang distribution \cite{belikov2017}.

\subsubsection{Lomax (Pareto) distribution}
Lomax distribution is also known as Pareto type II distribution \cite{lomax} and has the following probability density function 
\begin{equation}
    \rho_{lom}(z,\alpha)=\frac{\theta(z)\alpha}{\mu} \left[1+\frac{z}{\mu}\right]^{-(\alpha+1)},
    \label{pdf_lomax}
\end{equation} 
where $\mu>0$ is the scale parameter and $\alpha>1$ is the shape parameter. For $\alpha>1$ its mean is $\zeta=\mu/\left(\alpha-1\right)$ and for $\alpha>2$ its variance is $\sigma^2=\mu^2\alpha/\left(\left(\alpha-1\right)^2\left(\alpha-2\right)\right)$. For $\alpha \le 1$ this distribution has infinite mean and for $\alpha \le 2$ variance is infinite. It is essentially a Pareto probability density that has been shifted so that its support begins at zero. Note that for large $\alpha$ Lomax distribution tends to the exponential one. It represents a class of fat-tailed distributions which decays to zero as a power law. This feature should be contrasted e.g. with the log-normal distribution that is heavy-tailed but not fat-tailed meaning that it goes to zero slower than the exponential but faster than the power law. Lomax distribution has been widely used in economics as well as queueing theory \citep{johnson1995}.

In Fig. \ref{fig1} we compare the dimensionless probability densities $\rho(z)$ of the amplitudes $\{z_i\}$ of the Poissonian shot noise. The scaling is done via their mean $\zeta$ to make the densities $\rho(z)$ independent of the latter parameter and illustrates interrelations between them for the arbitrary but fixed $\zeta$.

\subsection{Quantity of interest: diffusion coefficient}
The most fundamental quantity characterizing the diffusive behaviour of the system is the diffusion coefficient given as
\begin{equation}
	D= \lim_{t \to \infty} \frac{\sigma_x^2(t)}{2t} = \lim_{t \to \infty} \frac{\langle x^2(t) \rangle - \langle x(t) \rangle^2}{2t},
\end{equation}
where $\sigma_x^2(t)$ is the variance of the particle coordinate $x(t)$. The averaging $\langle \cdot \rangle$ indicates 
\begin{equation}
	\langle x^k(t) \rangle = \int_{-\infty}^{\infty}  x^k P(x,t) dx.
\end{equation}
The probability density $P(x,t)$ corresponding to the particle position $x(t)$ fulfills the integro-differential master equation \cite{hanggi1978,hanggi1980}
\begin{align}
	\frac{\partial}{\partial t} P(x,t) = &-\frac{\partial}{\partial x}
[-U'(x) P(x,t)] + D_T \frac{\partial^2}{\partial x^2} P(x,t) \nonumber \\ &+ \lambda \int_{-\infty}^{\infty} [P(x-z,t) -P(x,t)] \rho(z) \,dz
	\label{master}
\end{align}
which can be reformulated as a spatially non-local diffusion equation, i.e.
\begin{align}
	\frac{\partial}{\partial t} P(x,t) = &-\frac{\partial}{\partial x}
[-U'(x) P(x,t)] \nonumber \\ &+ \frac{\partial^2}{\partial x^2} \int_{-\infty}^{\infty} \mathcal{D}(x,z) P(z,t) dz,
\end{align}
with an effective diffusion function
\begin{equation}
	\mathcal{D}(x,z) = D_P \rho(x - z) + D_T \delta(x - z)
\end{equation}
which consists of nonlocal (Poissonian) and local (thermal) parts \cite{czernik1997}.
\begin{figure}[t]
	\centering
    \includegraphics[width=0.9\linewidth]{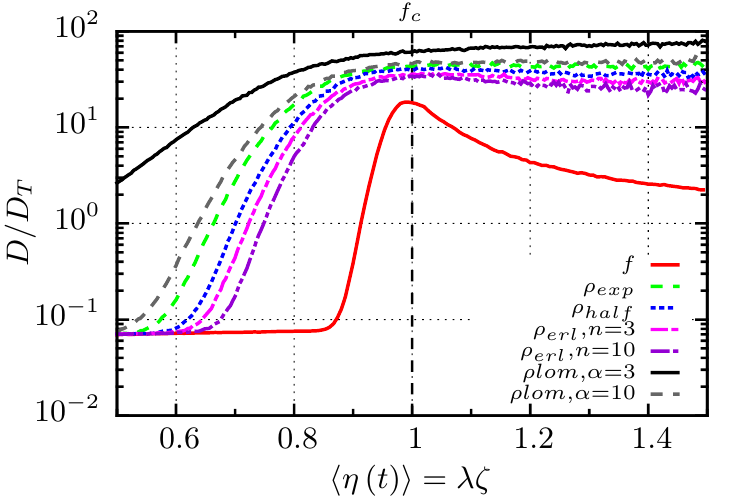}
    \caption{The rescaled diffusion $D/D_T$ where $D_T$ is the free diffusion coefficient versus the average value $\langle \eta(t) \rangle = \lambda \langle z_i \rangle = \lambda \zeta$ of the biased white Poissonian shot noise $\eta(t)$ presented for different amplitude $\{z_i\}$ statistics $\rho(z)$. The red solid line represents diffusive behaviour induced by the constant force $f(t) = f$. The spiking rate $\lambda = 10$ and thermal noise intensity $D_T = 0.01$ are fixed.}
    \label{fig2}
\end{figure}

\section{Results}
The diffusion coefficient $D$ can be calculated analytically for the system obeying Eq. (\ref{dimless_model}) with the external constant force $f(t) = f$. We refer the reader to Ref. [\onlinecite{reimann2001a}] and [\onlinecite{reimann2002}] for details of the calculations. The exact expression in the dimensional units reads
\begin{equation}
    D=D_0 \frac{\int_{x_0}^{x_0+L}\frac{dx}{L}I_{+}^2\left(x\right)I_{-}\left(x\right)}{\left[\int_{x_0}^{x_0+L}\frac{dx}{L}I_{+}\left(x\right)\right]^3},
\end{equation}
here $D_0$ is the Einstein free diffusion coefficient \mbox{$D_0:=k_B T/\Gamma$} ($\Gamma$ stands for the friction coefficient), $x_0$ is the arbitrary reference point and $I_{\pm}(x)$ is defined as
\begin{equation}
    I_{\pm}\left(x\right):=\int_{0}^{L}\frac{dy}{D_0}e^{\left\{\pm U\left(x\right)\mp U\left(x\mp y\right)-yF\right\}/k_B T}.
\end{equation}
In Ref. [\onlinecite{reimann2001a}] the authors reported that for weak thermal noise and near the critical tilt $f = f_c \approx 1$, the diffusion coefficient is gigantically enhanced versus the free diffusion. In such a case the dynamics given by Eq. (\ref{dimless_model}) can be divided into two processes, (i) the particle relaxation towards the minimum of the potential $U(x)$ as well as (ii) thermal noise driven escape from the latter position. The first time is robust with respect to temperature variation but the escape time is very sensitive to changes of this parameter. Such dichotomy lays at the root of the giant diffusion effect.
\begin{figure}[t]
	\centering
    \includegraphics[width=0.9\linewidth]{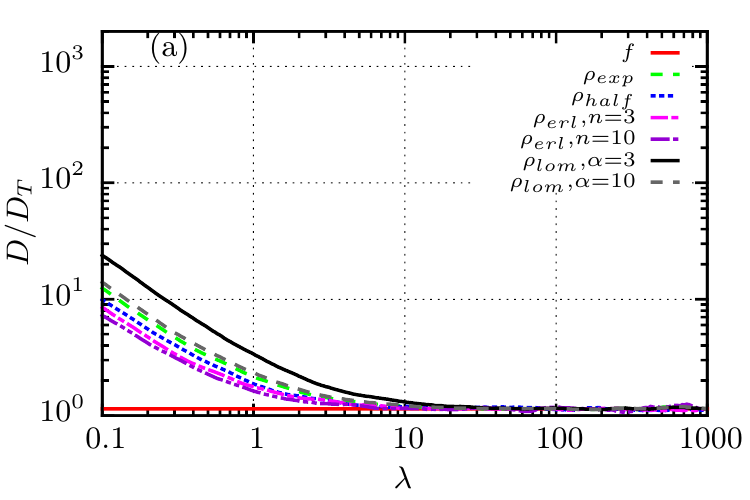}
    \includegraphics[width=0.9\linewidth]{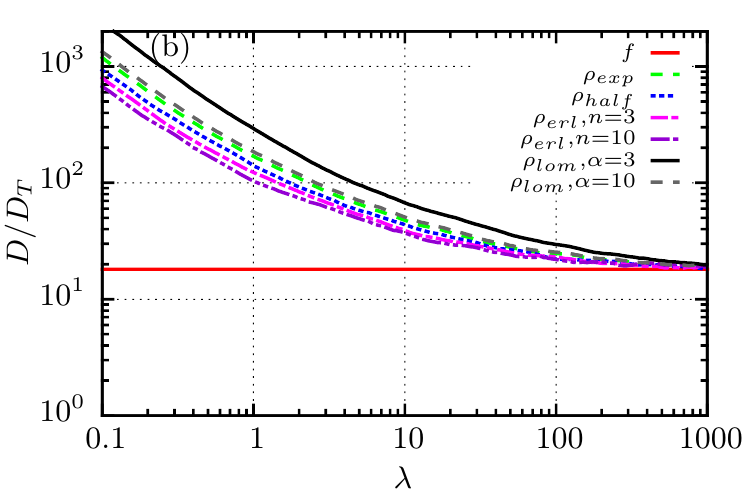}
    \caption{The rescaled diffusion coefficient $D/D_T$ presented as a function of the spiking rate $\lambda$ for different amplitude distributions $\rho(z)$ and thermal noise intensity (a) $D_T = 1$; (b) $D_T = 0.01$. The Poissonian shot noise mean is fixed $\langle \eta\left(t\right) \rangle= 1$. The red solid line corresponds to the system driven by the static force $f(t) = f$ whereas the other ones indicate influence of the nonequilibrium noise $\eta(t)$.}
    \label{fig3}
\end{figure}

Very recently this problem has been revisited in Ref. [\onlinecite{bialas}]. The authors demonstrated how the latter effect of giant diffusion is modified when the constant force $f(t) = f$ is replaced by nonequilibrium noise $f(t) = \eta(t)$ in the form of Poissonian shot noise. They reported a novel class of Brownian, yet non-Gaussian diffusion, in which the mean square displacement of the particle grows linearly with time and the probability density for the particle spreading is Gaussian, but the probability density for its position increments possesses an exponentially decaying tail. Moreover, the latter property leads to colossal enhancement of diffusion, pronouncedly exceeding the well known effect known as the giant diffusion.

In this paper we want to significantly expand these prediction by analyzing how the colossal diffusion is affected by a Poissonian shot noise amplitude statistics. The dynamics modeled by Eq. (\ref{dimless_model}) is described by the integro-differential master equation (\ref{master}) whose solution can not be solved in a analytical way. For this reason we study this issue by resorting to the comprehensive numerical simulations.
\begin{figure}[t]
	\centering
    \includegraphics[width=0.9\linewidth]{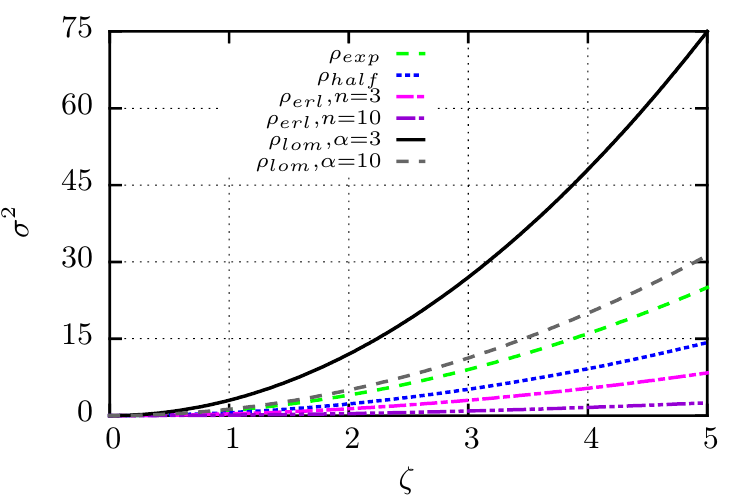}
    \caption{The variance $\sigma^2$ of amplitude distributions $\rho(z)$ presented as a function of their mean $\zeta$.} 
    \label{fig4}
\end{figure}

All numerical calculations have been done by the use of a Compute Unified Device Architecture (CUDA) environment implemented on a modern desktop Graphics Processing Unit (GPU). This proceeding allowed for a speedup of factor of the order $10^3$ times as compared to present day Central Processing Unit (CPU) method \cite{spiechowicz2015cpc}. Unless stated otherwise, the quantities of interest characterizing diffusive behaviour of the system were averaged over Gaussian $\xi(t)$ and Poissonian $\eta(t)$ noise realizations forming the ensemble of $2^{16} = 65536$ trajectories, each starting with different initial condition $x(0)$ distributed uniformly over the spatial period $[0, 2\pi]$ of the potential $U(x)$.
\begin{figure*}[t]
	\centering
	\includegraphics[width=0.45\linewidth]{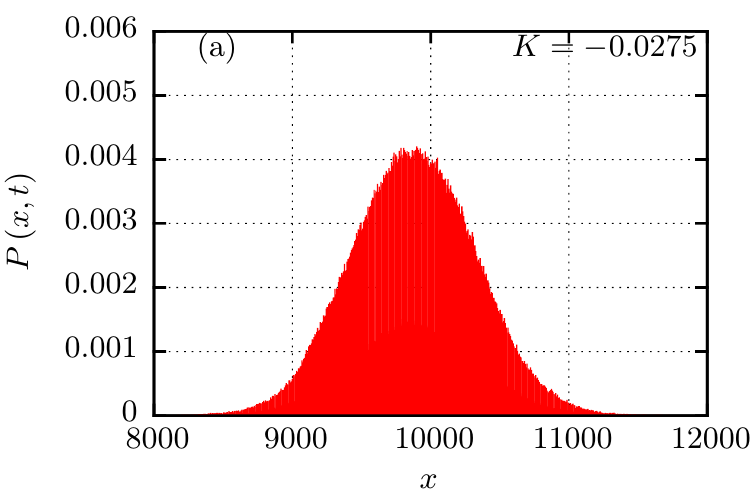}
	\includegraphics[width=0.45\linewidth]{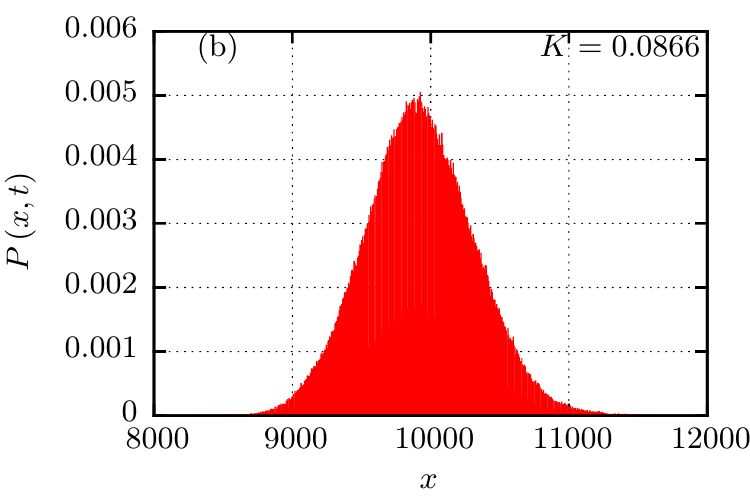}\\
	\includegraphics[width=0.45\linewidth]{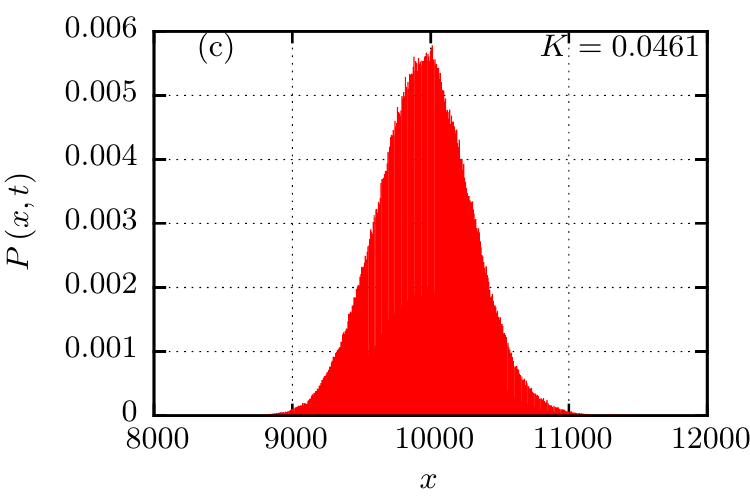}
	\includegraphics[width=0.45\linewidth]{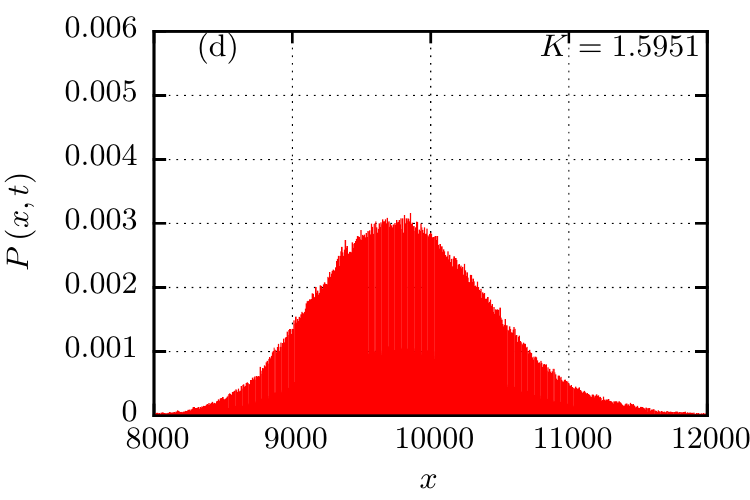}\\
	\caption{The probability distribution $P(x,t)$ of the particle coordinate $x$ at time $t$ for the following nonequilibrium noise $\eta(t)$ amplitude statistics $\rho(z)$: (a) exponential $\rho_{exp}(z)$, (b) half-normal $\rho_{half}(z)$, (c) the Erlang $\rho_{erl}(z,n)$ with $n=10$ and (d) the Lomax $\rho_{lom}(z,\alpha)$ with $\alpha=3$. Other parameters are thermal noise intensity $D_T=0.01$, the spiking rate $\lambda=0.1$ and the mean amplitude $\zeta=10$ (i.e. $\langle \eta(t) \rangle=1$). The above distributions were calculated for $t = 10 000$ for which we checked that diffusion is normal $\sigma^2_x(t)\sim Dt$.}
	\label{fig5}
\end{figure*}

\subsection{Colossal diffusion}
In Fig. \ref{fig2} we illustrate the rescaled diffusion coefficient $D/D_T$ where $D_T$ is the free thermal diffusion versus the averaged value $\langle \eta(t) \rangle = \lambda \langle z_i \rangle = \lambda \zeta$ of the biased white Poissonian noise $\eta(t)$ presented for different amplitude $\{z_i\}$ statistics $\rho(z)$. The spiking rate $\lambda = 10$ and thermal noise intensity $D_T = 0.01$ are fixed. The red solid line corresponds to the diffusive behaviour induced by the static force $f(t) = f$. The first observation is that the latter perturbation indeed enhances the diffusion coefficient over the free thermal diffusion. This effect is most pronounced near the critical tilt $f = f_c = 1$. Secondly, the Poissonian shot noise amplifies the diffusion coefficient much more pronouncedly and the latter feature is detected not only near the critical region $\langle \eta(t) \rangle = f_c = 1$ but for a significantly wider window of the average bias $\langle \eta(t) \rangle$. The impact of nonequilibrium noise amplitude statistics is also visualized in Fig. \ref{fig1}. The general observation is that the distributions for which the decay rate is slower enhance the diffusion coefficient in a greater extent. It is clearly visible in the case of the Lomax (Pareto) density which displays the algebraic tail and as a consequence maximize the colossal diffusion amplification.

In Fig. \ref{fig3} we present the diffusion enhancement $D/D_T$ as a function of the spiking rate $\lambda$ for the fixed mean $\langle \eta(t) \rangle = f_c = 1$ and different amplitude distributions $\rho(z)$. Subplots (a) and (b) correspond to thermal noise intensity $D_T = 1$ and $D_T = 0.01$, respectively. The red solid line represents giant diffusion induced by the static bias $f$ whereas the dashed lines indicates the influence of the Poissonian shot noise. In the limiting case of large $\lambda \to \infty$ and small $\zeta \to 0$ (remember that the mean $\langle \eta(t) \rangle = \lambda \zeta$ is fixed), namely, for very frequent $\delta$-kicks of tiny amplitude the rescaled diffusion coefficient $D/D_T$ tends to the value characteristic for the system driven by the constant force \cite{spiechowicz2014pre}. This observation holds true regardless of temperature, c.f. panel (a) vs (b), and irrespective of the Poissonian shot noise amplitude statistics. Similarly, in the limit $\lambda \to 0$ and $\zeta \to \infty$, i.e. when the particle is rarely kicked by very strong $\delta$-kicks the rescaled diffusion coefficient $D/D_T$ is much larger than it is the case for the deterministic force $f$ thus indicating the colossal, instead of giant diffusion \cite{bialas}. When temperature $D_T$ increases the colossal diffusion, i.e. amplification of diffusion over the giant one observed for the static force $f$, is detected for smaller spiking rates $\lambda$ of the nonequilibrium noise $\eta(t)$ and the magnitude of enhancement is smaller as well.

Moreover, we note that the hierarchy of the amplification induced by different amplitude statistics $\rho(z)$ observed in Fig. \ref{fig2} is robust with respect to alteration of the kicks frequency $\lambda$ as well as temperature $D_T$. It can be ordered from the smallest to the largest as follows: the Erlang $\rho_{erl}$, half-normal $\rho_{half}$, exponential $\rho_{exp}$ and the Lomax (Pareto) $\rho_{lom}$ statistics. Such a hierarchy can also be deduced in distributions ranked in terms of the rate of their decay from the fastest to the slowest. This characteristics is shown in the inset of Fig. \ref{fig1}. We again conclude that the distributions for which the decay rate is slow enhance the diffusion coefficient in a greater extent. We quantify the latter aspect in Fig. \ref{fig4} where we present the variance $\sigma^2$ of amplitude distributions $\rho(z)$ presented as a function of their mean $\zeta$. The reader can notice that the hierarchy of the curves presented there directly corresponds to the ordering observed in Fig. \ref{fig3}. Therefore variance $\sigma^2$ of the amplitude distribution can be used to determine the relation between the enhancement induced by different statistics. In the limit of large spiking rate $\lambda \to \infty$ and tiny amplitude $\zeta \to 0$ the mentioned hierarchy disappears. Then the rescaled diffusion coefficient $D/D_T$ tends to the value characteristic for the system driven by the constant force and the difference between various amplitude statistics becomes negligible. On the other hand, in the limit of small spiking rate $\lambda \to 0$ and huge amplitude $\zeta \to \infty$, in which the colossal diffusion is observed, we found that the ratio between rescaled diffusion coefficients $D/D_T$ for different statistics $\rho(z)$ is constant and can be well approximated by the proportion of their variance $\sigma^2$ (not depicted). The expressions for the mean and variance of the distributions $\rho(z)$ are of the form $\zeta = c_1 \mu$ and $\sigma^2 = c_2 \mu^2$, where $c_1$ and $c_2$ are constants. After a quick transformation we obtain $\sigma^2 = \zeta^2 c_2/c_1^2$. For the settled $\lambda$ (and therefore fixed $\zeta$ due to the constraint $\langle \eta(t) \rangle = \lambda \zeta = 1$) the ratio of variance is constant as well. This property leads to the fixed proportion of diffusion coefficients $D/D_T$ for different amplitude statistics $\rho(z)$ in the limit where Poissonian shot noise dominates the dynamics, c.f. Fig. \ref{fig3}.

\begin{figure}[t]
	\centering
	\includegraphics[width=1.0\linewidth]{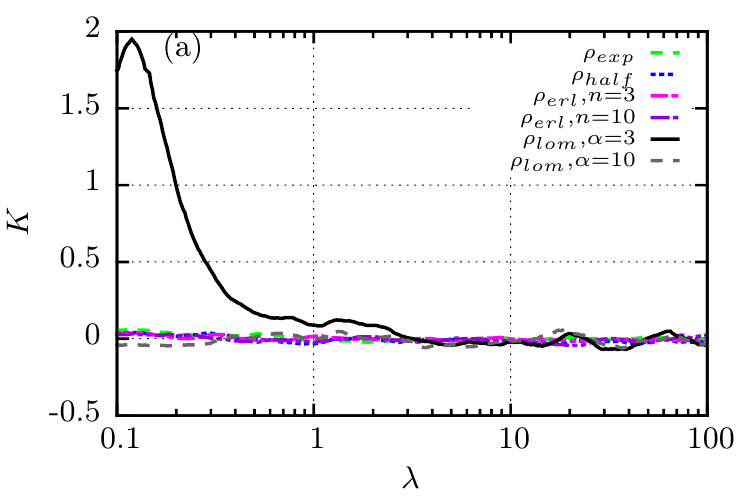}\\
	\includegraphics[width=1.0\linewidth]{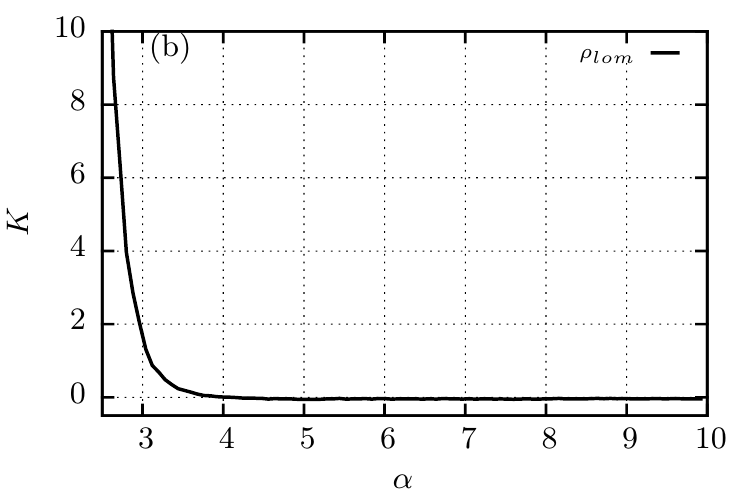}
	\caption{Panel (a): the kurtosis $K$ of the probability distribution $P(x,t)$ of the particle coordinate $x$ at time $t$ versus the spiking rate $\lambda$ presented for different amplitude statistics $\rho(z)$ of the nonequilibrium noise $\eta(t)$. Thermal noise intensity reads $D_T = 0.01$. Panel (b): the kurtosis $K$ for the Lomax amplitude distribution $\rho_{lom}(z,\alpha)$ as a function of the power exponent $\alpha$. The spiking rate and temperature reads $\lambda = 0.1$ and $D_T = 0.01$, respectively. In both panels $K(t)$ was determined at $t = 10 000$ for which diffusion is already normal.}
	\label{kurt}
\end{figure}
\subsection{Brownian, yet non-Gaussian diffusion}
The spread of trajectories of the system given by Eq. (\ref{dimless_model}) described by the mean square displacement of the particle coordinate scales linearly with time $\sigma_x^2(t) \sim D t$ in an asymptotic regime. It means that the observed diffusion is normal (Brownian). However, nowadays it is clear that the latter does not necessarily imply the Gaussianity of the probability density for the particle coordinate. Therefore in Fig. \ref{fig5} we present the probability distributions $P(x,t)$ of the particle position $x$ at time $t$ for various nonequilibrium noise $\eta(t)$ amplitude statistics $\rho(z)$ depicted for the spiking rate $\lambda = 0.1$ and the mean amplitude $\zeta = 10$, i.e. $\langle \eta(t) \rangle = 1$. Their Gaussianity can be quantified by the kurtosis $K(t)$ defined as
\begin{equation}
K(t) = \frac{\left \langle [x(t) - \langle x(t) \rangle]^4 \right \rangle}{\left\{ \left\langle [x(t) - \langle x(t) \rangle]^2 \right \rangle \right\}^2} - 3 \;,
\end{equation}
For the Gaussian density this quantity assumes zero, i.e. $K(t) = 0$. In the studied case, $K(t)$ was calculated at $t = 10 000$ for which the mean square displacement of the particle scales linearly with time and therefore the diffusion is already normal (Brownian). We find that for the amplitude statistics whose tails are exponentially bounded, i.e. not heavy-tailed, the kurtosis yields approximately zero $K(t) \approx 0$, indicating that the probability distribution $P(x,t)$ of the particle coordinate is Gaussian. 
On the other hand, for the case of the Lomax amplitude statistics (which is fat-tailed) with $\alpha = 3$ the kurtosis $K(t) = 1.5951$ significantly differs from zero leading to the Brownian (normal) yet non-Gaussian diffusion.

In Fig. \ref{kurt} (a) we illustrate how the kurtosis $K(t)$ of the probability distribution $P(x,t)$ computed at $t = 10 000$ depends on the spiking rate $\lambda$ for different amplitude distributions $\rho(z)$ of the nonequilibrium noise $\eta(t)$. The main observation coming from the inspection of this panel is that for all statistics $\rho(z)$ considered in this work except of the Lomax one the kurtosis is negligibly small $K(t) \approx 0$ and robust with respect to alteration of the spiking rate $\lambda$. However, for the Lomax amplitude distribution $\rho_{lom}(z,a)$ with small $\alpha$ the kurtosis $K(t)$ noticeably deviates from zero as the frequency of $\delta$-kicks vanishes $\lambda \to 0$.
\begin{figure*}
	\centering
	\includegraphics[width=0.45\linewidth]{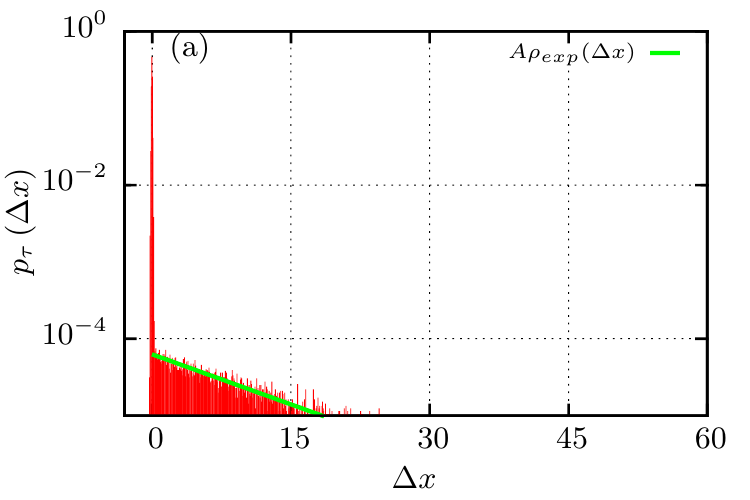}
	\includegraphics[width=0.45\linewidth]{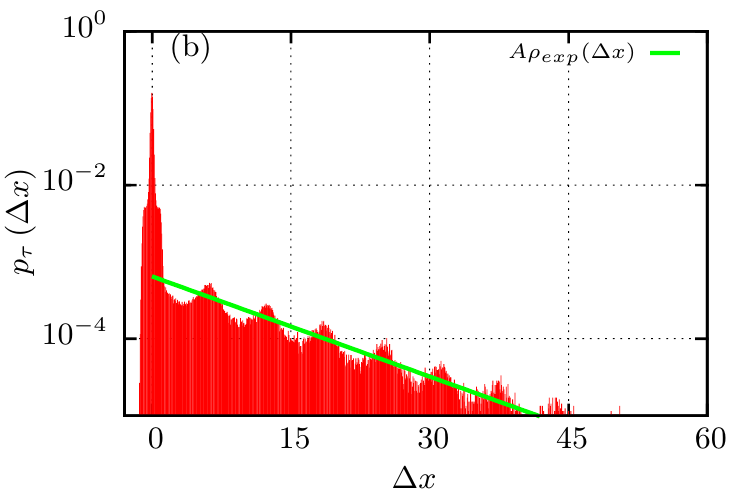}\\
	\includegraphics[width=0.45\linewidth]{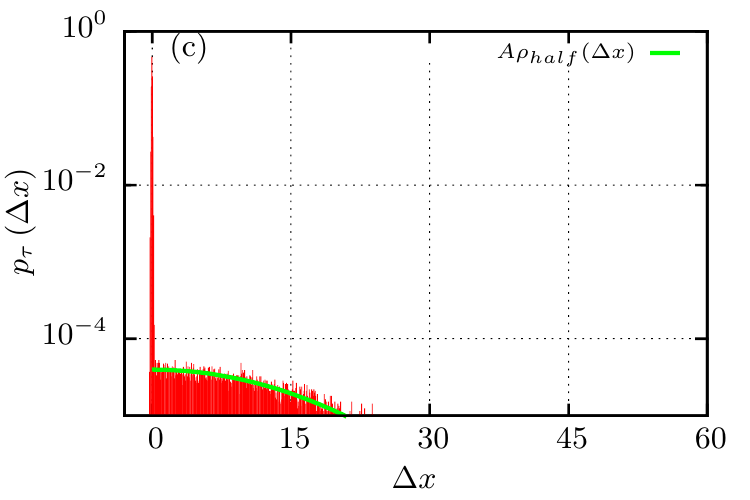}
	\includegraphics[width=0.45\linewidth]{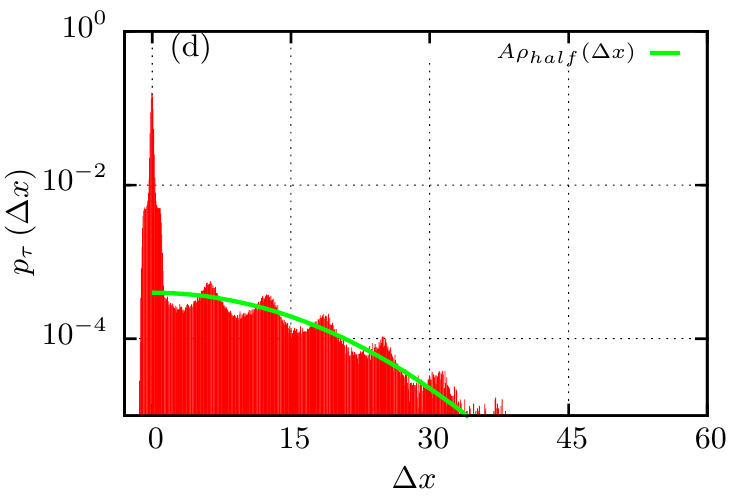}\\
	\includegraphics[width=0.45\linewidth]{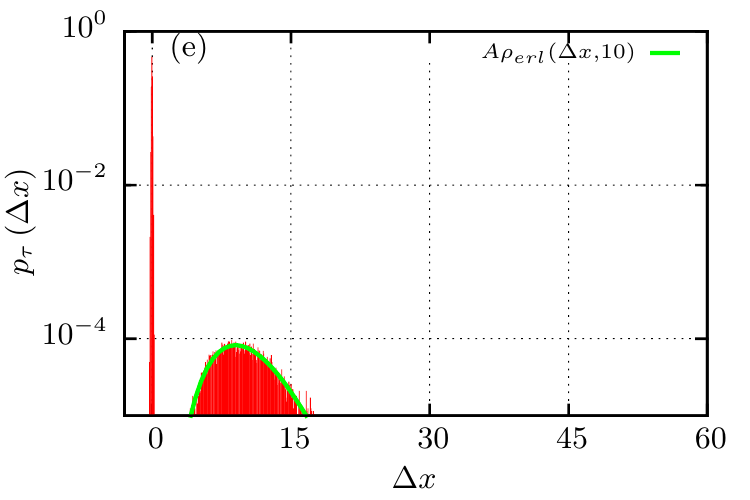}
	\includegraphics[width=0.45\linewidth]{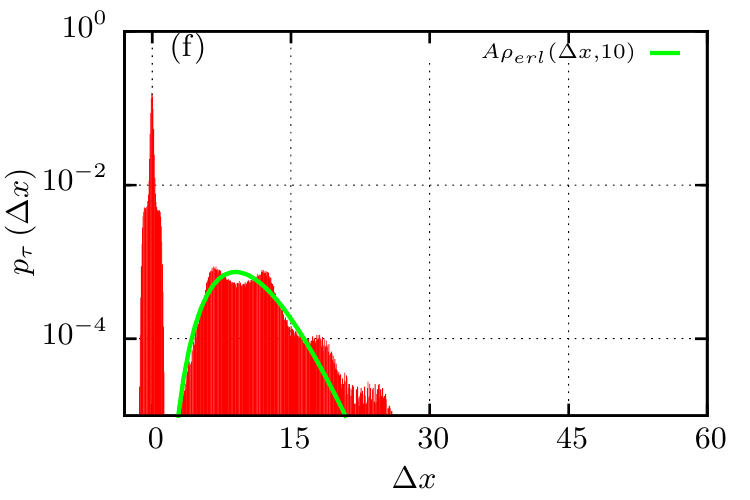}\\
	\includegraphics[width=0.45\linewidth]{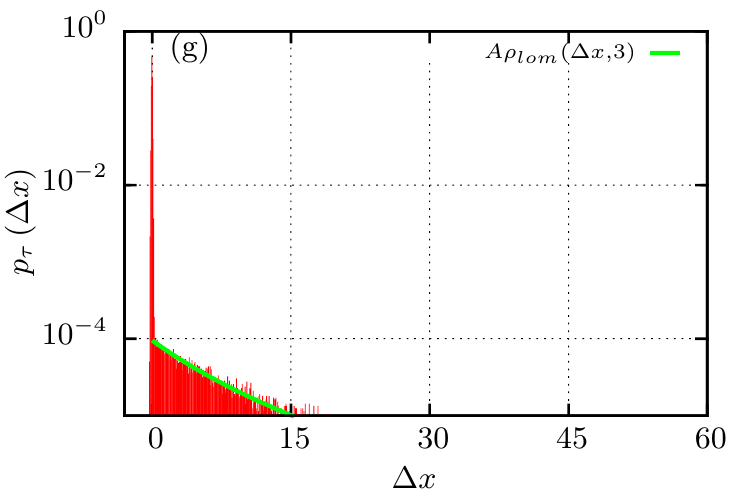}
	\includegraphics[width=0.45\linewidth]{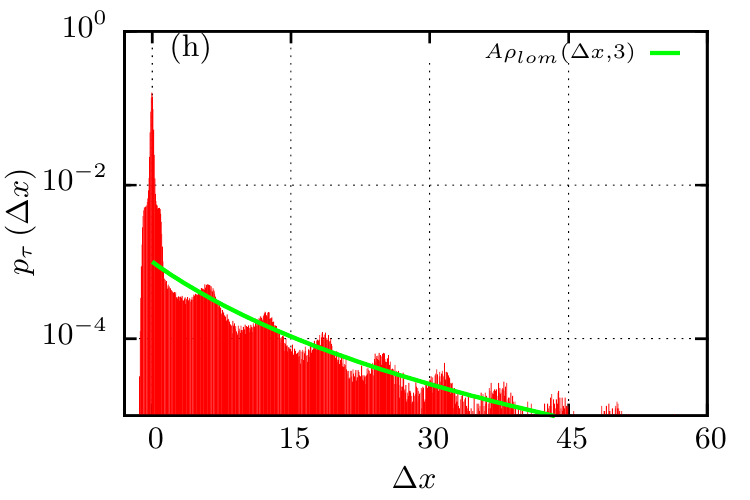}\\
	\caption{The probability distribution $p_{\tau}\left(\Delta x\right)$ for the long time particle position increments $\Delta x(\tau) = \lim_{t \to \infty}[x(t + \tau) - x(t)]$ depicted for different amplitude statistics $\rho(z)$. The left column corresponds to the time difference $\tau=0.1$ while the right to $\tau=1$. Panels (a) and (b): the exponential amplitude statistics $\rho_{exp}(z)$; (c) and (d): the half-normal $\rho_{half}(z)$; (e) and (f): the Erlang $\rho_{erl}(z,n)$ with $n=10$; (g) and (h): the Lomax statistics $\rho_{lom}(z,\alpha)$ with $\alpha=3$. Other parameters are as follows: thermal noise intensity $D_T=0.01$, the spiking rate $\lambda=0.1$ and the mean amplitude $\zeta=10$ (i.e. $\langle \eta\left(t\right) \rangle=1$). Fits of the corresponding distributions with the parameter $A$ are denoted by the green lines.}
	\label{fig6}
\end{figure*} 

This behaviour is related to Kolmogorov-Gnedenko generalized central limit theorem \cite{gnedenko} which states that the sum of a number of random variables with a power-law (Paretian) tail $|x|^{-(\alpha+1)}$ with $0 < \alpha < 2$ (infinite variance) will tend to a stable distribution \emph{as the number of summands grows}. For $\alpha > 2$ the sum converges to a Gaussian distribution. However, if the variance just barely exists, like it is in the case of $\alpha = 3$, the generalized central limit theorem in principle applies but may lead to a very bad approximation. The Berry-Esseen theorem \cite{feller1970} specifies the rate at which the convergence between the given distribution and a Gaussian density takes place. In the studied case it is slow. Since the time between successive Poisson arrival times is exponentially distributed we note that for the fixed spiking rate $\lambda$ the number of summands ($\delta$-kicks) grows when the time scale of the observation is increased. Therefore for $\alpha > 2$ the deviation of the kurtosis $K(t)$ from zero may be also interpreted as a finite size effect of the observation time scale.

We quantified these aspects in a deeper way in Fig. \ref{kurt} (b) where we depict the kurtosis $K(t)$ of the probability distribution $P(x,t)$ for the Lomax amplitude statistics $\rho_{lom}(z,\alpha)$ of the nonequilibrium noise $\eta(t)$ as a function of the power exponent $\alpha$. The spiking rate is fixed \mbox{$\lambda = 0.1$}. We can note that for a finite time scale $t = 10 000$ the kurtosis diverges $K(t) \to \infty$ when $\alpha \to 0$. On the other hand, for such a time span if $\alpha > 4$ then the kurtosis is negligible $K(t) \approx 0$. Therefore for the Lomax amplitude distribution $\rho_{lom}$ the power exponent $\alpha$ as well as the time scale is crucial to determine the Gaussianity of the particle coordinate probability distribution $P(x,t)$.

One of the fundamental features of diffusion (Wiener) process is Gaussianity of the probability density $P(x,t)$ for finding the particle at position $x$ at time $t$. However, it is a consequence of a defining property telling that its increments $\Delta x$ are distributed according to the Gaussian statistics $p(\Delta x)$ as well. Therefore as the next step we consider the probability density $p_{\tau}(\Delta x)$ of the particle position increments
\begin{equation}
	\Delta x(\tau) = \lim_{t \to \infty}[x(t + \tau) - x(t)],
\end{equation}
where $\tau$ is the time increment. In Fig. \ref{fig6} we present this quantity for different nonequilibrium noise $\eta(t)$ amplitude statistics $\rho(z)$ as well as two time lags $\tau$. The probability distribution $p_{\tau}(\Delta x)$ of the particle position increments visible there consists of two parts. The first one is associated with thermal equilibrium fluctuations. It is manifested as the well pronounced peak around $\Delta x = 0$. The second one is related to nonequilibrium fluctuations in the form of Poissonian shot noise. As it is depicted in Fig. \ref{fig6}, the latter is responsible for the tail of the distribution $p_{\tau}(\Delta x)$. In Ref. [\onlinecite{bialas}] the authors discovered that for the exponential amplitude statistics $\rho_{exp}(z)$ the probability distribution of the particle position increments $p_{\tau}(\Delta x)$ possesses an exponential tail. Here we generalize this result with observation that the tail of the density $p_{\tau}(\Delta x)$ is stemmed from the amplitude statistics $\rho(z)$, c.f. fits of the corresponding distributions marked with the green lines. Moreover, we find that for all considered cases the particle position increment distribution $p_{\tau}(\Delta x)$ is distinctly non-Gaussian. Therefore, for a broad spectrum of Poissonian shot noise amplitude statistics the dynamics given by Eq. (\ref{dimless_model}) describes Brownian (normal), yet non-Gaussian diffusion. The impact of the time lag $\tau$ on $p_{\tau}(\Delta x)$ can be summarized as follows. For increasing $\tau$ the cutoff of the distribution $p_{\tau}(\Delta x)$ grows. In the latter case the multi-peaked, comb-like structure of the statistics $p_{\tau}(\Delta x)$ is detected. It is characteristic for an overdamped dynamics in a periodic potential in which the particle quickly relaxes towards the neighboring potential minima.
\begin{figure}[t]
	\centering
	\includegraphics[width=1.0\linewidth]{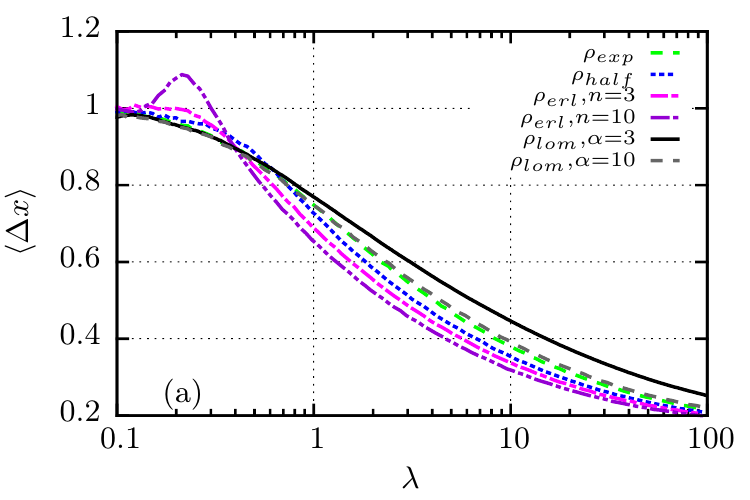}\\
	\includegraphics[width=1.0\linewidth]{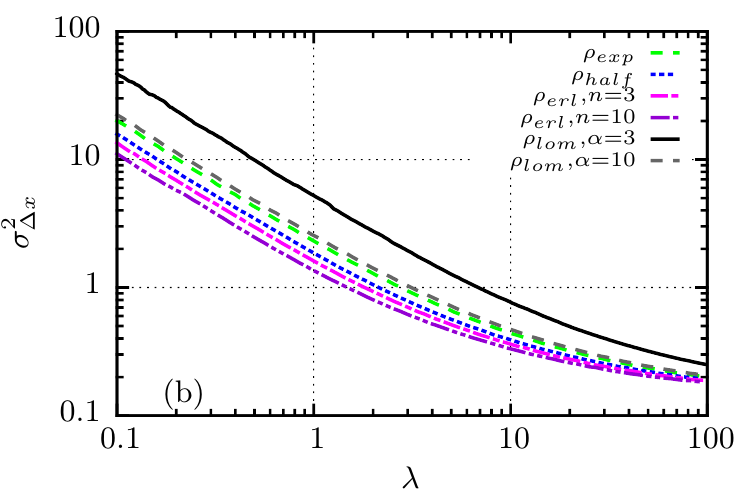}
	\caption{The mean $\langle \Delta x \rangle$ and the variance $\sigma^2_{\Delta x}$ of the particle position increment is depicted as a function of the spiking rate $\lambda$ for different amplitude statistics $\rho(z)$ of the nonequilibrium noise $\eta(t)$ in the panel (a) and (b), respectively. The time increments reads $\tau = 1$ and thermal noise intensity \mbox{$D_T = 0.01$.}}
	\label{inc}
\end{figure}

In Fig. \ref{inc} (a) we present the mean value $\langle \Delta x \rangle$ of the particle coordinate increment versus the spiking rate $\lambda$ for different amplitude statistics $\rho(z)$. The time lag is fixed $\tau = 1$. This characteristics can reflect the non-monotonic behaviour of the amplitude distribution $\rho(z)$ as it is the case for the Erlang statistics $\rho_{erl}$ with large $n$. In other cases the mean value $\langle \Delta x \rangle$ is a monotonically decreasing function of the spiking frequency $\lambda$. It is interesting to note that the hierarchy of curves depicted in the panel can be contrasted in two regimes. In the first one, which corresponds to the large spiking rate $\lambda \to \infty$ it is the same as the ordering observed for the diffusion amplification, see Fig. \ref{fig3} (b). It means that the mean value $\langle \Delta x \rangle$ of the increment is larger if the decay rate of the amplitude statistics $\rho(z)$ is slower. On the other hand, when the spiking rate is small $\lambda \to 0$ the hierarchy of curves is reversed, i.e. the average increment $\langle \Delta x \rangle$ grows if the decay rate is faster. In the panel (b) we illustrate the variance of the particle position increment $\sigma^2_{\Delta x} = \langle (\Delta x)^2 \rangle - \langle \Delta x \rangle^2$ versus the frequency $\lambda$ for the same time lag $\tau = 1$. Regardless of the amplitude statistics $\rho(z)$ this quantity is a monotonically decreasing function of the spiking rate $\lambda$ as it is the case for the rescaled diffusion coefficient $D/D_T$, see Fig. \ref{fig3}.

\section{Conclusions}
In this paper we considered a paradigmatic model of nonequilibrium statistical physics consisting of the overdamped Brownian particle dwelling in a periodic potential. When the system is driven by nonequilibrium noise the particle diffusion coefficient $D$ may be colossally enhanced, extremely exceeding the previously studied situation known as the giant diffusion. Such a scenario is vital for a correct description of biophysical systems which immanently operate under nonequilibrium conditions and are exposed to non-thermal perturbations.

As a model of nonequilibrium stochastic force we investigated the Poissonian white shot noise $\eta(t)$. We analyzed how the colossal diffusion phenomenon is affected by the Poissonian shot noise amplitude statistics $\rho(z)$. We considered a broad spectrum of different probability distributions that allows us to study the influence of non-monotonicity as well as exponential, superexponential (e.g. Gaussian) and subexponential (algebraic) decay in the density. It turned out that the tail of amplitude distribution has crucial impact on the magnitude of colossal diffusion amplification. The general observation is that the distributions for which the decay rate is slower enhance the diffusion coefficient in a greater extent. We quantified this aspect by the variance of distribution $\sigma^2$ which may serve as a convenient tool to evaluate interrelations between the magnitude of diffusion enhancement. If for two amplitude densities $\sigma^2_2 > \sigma^2_1$, then the corresponding diffusion coefficients $D_2 > D_1$.

The origin of the colossal amplification of diffusion is rooted in existence of non-Gaussian statistics in the particle coordinate increments $\Delta x$. This fact should be contrasted with conventional Brownian motion for which the latter are distributed according to the Gaussian distribution and as a consequence the probability density $P(x,t)$ for finding the particle at position $x$ at time $t$ is also Gaussian. We found that for all considered cases the mean square displacement of the particle grows linearly with time $\sigma_x^2 \sim D t$, however, 
the particle position increment distribution $p(\Delta x)$ is distinctly non-Gaussian and its tail is originated from the nonequilibrium noise amplitude density $\rho(z)$. Therefore for this broad class of statistics Poissonian shot noise induces Brownian (normal), yet non-Gaussian diffusion. Moreover, it turned out that if the nonequilibrium noise amplitude density is exponentially bounded then the distribution $P(x,t)$ is Gaussian while maintaining $p(\Delta x)$ non-Gaussian. In contrast, for a fat-tailed amplitude statistics $\rho(z)$ both distributions $P(x,t)$ and $p(\Delta x)$ can be non-Gaussian.

Our findings for the paradigmatic model of nonequilibrium statistical physics can be applied to numerous systems \cite{lee2006,reimann2008,ma2015,risken} including cold atoms dwelling in optical lattices \cite{lutz2013,kindermann2017,dechant2019} as well as complex fluids \cite{song2019,miotto2021} and therefore we think the results will inspire a vibrant followup of both experimental and theoretical studies. In addition, they carry profound consequences for a wide spectrum of first arrival problems \cite{lanoiselee2018}.
\section*{Acknowledgment}
This work has been supported by the Grant NCN No. 2017/26/D/ST2/00543 (J. S.).

\section*{Conflict of interest}
The authors have no conflicts to disclose.

\section*{Data availability statement}
The data that support the findings of this study are available from the corresponding author upon reasonable request.

\end{document}